# A Review of The Algebraic Approaches to Quantum Mechanics. Appraisals on Their Theoretical Relevance

Antonino Drago
Formerly at University "Federico II" Naples – I – drago@unina.it

**Abstract:** I review the various algebraic foundations of quantum mechanics. They have been suggested since the birth of this theory till up to last year. They are the following ones: Heisenberg-Born-Jordan's (1925), Weyl's (1928), Dirac's (1930), von Neumann's (1936), Segal's (1947), T.F. Jordan's (1986), Morchio and Strocchi's (2009) and Buchholz and Fregenhagen's (2019). Three cases are stressed: 1) the misinterpretation of Dirac's foundation; 2) von Neumann's 'conversion' from the analytic approach of Hilbert space to the algebraic approach of the rings of operators; 3) the recent foundation of quantum mechanics upon the algebra of perturbation Lagrangians. Moreover, historical considerations on the go-and-stop path performed by the algebraic approach in the history of QM are offered. The level of formalism has increased from the mere introduction of matrices till up to group theory and C*-algebras. But there was no progress in approaching closer the foundations of physics; therefore the problem of discovering an algebraic formulation of QM organized as a problem-based theory and making use of no more than constructive mathematics is open.

## 1. Introduction

More than Planck's results on black-body, Einstein's 1905 paper on quanta introduced the discrete into theoretical physics. Its first section ("Introduction") contrasts the discrete and the continuum inside classical theoretical physics. Afterwards in theoretical physics the discrete was unavoidable and took a place at the same par of the continuum in the mathematical representation of reality. At the birth of quantum mechanics (QM) in the year 1925, the invention of Matrix mechanics - essentially based on the discrete and algebra – was quickly followed by the invention of Wave mechanics - essentially based on the continuum and differential equations -, The usual presentation of QM is obliged to contrast discrete with continuum.

However, after the unification in the same year of them and mainly after Dirac and von Neumann proofs of their equivalence, this dichotomy discrete/continuum appeared as overcome not only in the mathematical formalism but also at the theoretical level. However some scholars insisted in recognize a dichotomy, albeit of a philosophical nature, concerning two different theoretical approaches to QM. The most relevant one came from Fred Kronz and Tracy Lupher (2019); they contrasted the "rigor" of von Neumann's approach with the "pragmatism" of Dirac's approach.

Rather one may assume the old mathematical dichotomy, that between analysis and algebra, which generalizes the dichotomy discrete/continuum and even more the above philosophical one.

The present paper reviews the introduction of the algebraic approach into the physical theory of QM, which more than special relativity gave a new address to theoretical physics.

The algebraic approach in theoretical physics started in the 19$^{th}$ century through crystallography – a branch of science originating group theory.. Yet, a further introduction of this kind of approach required much time. Also Albert Einstein who introduced Lorentz group of space-time transformations of electromagnetism into mechanics did not mention its algebraic name (the word "group" occurred in the generic sense of a gathering). Instead, in QM an algebraic approach started since the origin of this theory through Heisenberg's formulation. Yet, later the analytical approach of differential equations and then Hilbert space prevaile1d and became almost the unique approach. However, it is interesting that in subsequent times the algebraic approach slowly but decisively gained importance in the foundations of QM. In the following eight different foundations of QM according to the algebraic approach will be presented.

Even more general than the dichotomy analysis/algebra is the dichotomy on the two basic kinds of mathematics, the constructive mathematics, born in the '60s under the requirement of constructing all mathematical objects,(Markov 1962, Bishop 1967) and the classical one. This dichotomy is marked by the rejection or not of the idealistic notions: e.g. Weierstrass theorem on an accumulation point of an infinite set of points, Heine-Borel theorem, Dirac's $\delta(x)$, Hilbert space, etc. A quick review of the basic choices on which each algebraic foundation relies is suggested. No algebraic full formulation of QM relying on the alternative choices to that of von Neumann's formulation was discovered.

## 2. Heisenberg's foundation of QM (1925)

Werner Heisenberg's foundation represents the first attempt of formulating QM. Being aimed at solving in strict operative terms the problem of spectra through the basic concept of the electromagnetic atom it made use of an algebraic mathematics without making a priori discreteness assumption, as instead Max Born then did by translating Heisenberg's calculations into the little known mathematical objects, algebraic matrices; then Pascual Jordan improved this matricial calculus.(Born, Heisenberg, Jordan 1925) The representation of physical operators by matrices incorporates a dependence on time, whereas that of states-vectors is time independent. The main mathematical problem is the resolution of the eigenvalue problem of a matrix in order to obtain the values of the corresponding observable. This theory was the first capable to explain in a theoretical way basic quantum phenomena (one-dimensional harmonic oscillator, rotator, Zeeman effect and the fluctuations of the radiation field). However, one year after it was proven that this theory is equivalent to, and less easy than the next foundation of quantum theory, wave mechanics where operators are constant and the states are time dependent. In the subsequent times H-B-J foundation substantially changed.(Beller 1983, p. 470) At present these two foundations of QM are recalled - by leaving aside that they naturally descend from the classical Hamiltonian (Donini 1983; Darrigol 2014, pp. 248-253) - as the two main representations of von Neumann's formulation of QM in the infinite dimensional Hilbert space. No explanation is offered by historians why this first algebraic foundation could then be included into a more general formulation cancelling its algebraic approach; it is probable that the effectiveness in solving physical problems (also of foundational nature) through the use of Hilbert space prevailed so much on the worries to understand the foundations of QM to make equivalent two mathematical approaches, the algebraic one and the analytical one, whose natures are mutually antagonistic. Mainly Norwood R. Hanson (1963), F.A. Muller (1997) and Enrico A. Giannetto (1997) contested this unification. The last one lucidly stressed that the unification disregarded exactly this point, i.e, the different kinds of relationship mathematics-physics of the two formulations.(p. 203)

> Matrix mechanics implied a change in the relationship between mathematics and physics: as in the case of relativity which involved changing geometry and assuming a "physical geometry" chosen in agreement of its experimentability, so arithmetic and algebra in the case of matrix mechanics could no longer be given a priori, they had to be chosen in relation to experiments, and this led to a "quantum" arithmetic and algebra or "quantum numbers" (matrices of "q-numbers"). The revolutionary consequence of this is the overturning of the relations between "logos mathematikòs" and " physis", which will then be carried out by the radical approaches of "quantum logic". To analyze still today the constitutive works of the mechanics of matrices then does not only have the flavor of a nostalgia or a historical erudition; instead, it concerns a restoration of the extreme radicalism of the original "quantum revolution", not only for its historical value, but for its physical and philosophical consequences that oblige us to completely reconsider our own way to relate us to nature from both an epistemological point of view and a point of view that cannot fail to be ethical too: the change in the idea of nature, implied by the mechanics of matrices, involves the abandonment of the illusory modern "era of the images of the world", in which nature was reduced to an image of man … (Giannetto 1997, p. 203)

It is also interesting that after the birth of QM a great discussion led physicists to recognize the problem of the simultaneous measurements of two physical quantities; at last Heisenberg introduced the so-called "uncertainty principle", which states that position $Q$ and momentum $P$ cannot be measured simultaneously with an unlimited accuracy. (Heisenberg 1927). Surely it does not represent an axiom for a deductive system, rather a methodological principle for solving the basic problem of QM of recuperating the measurement of a physical system. Born's idea of representing the variables through matrices resulted a wise move, because their multiplication is non-commutative and hence this non-classical situation was represented through the algebraic language: $QP-PQ = ih/2\pi$, where P and Q are matrices. It may be called the "strange equation".(Jordan 1986, p. 5) because it is at all surprising a classical physicist.

## 3. Weyl's book *Group Theory and Quantum Mechanics* (1928)

George W. Mackey summarizes Weyl's novelties in quantum physics:

> … the contributions of Hermann Weyl… began in 1924 with the work on extending the representation theory of finite groups to a class of infinite continuous groups – the compact Lie groups. They culminated in 1927 with Weyl's work in: (a) Unifying group representation theory with Fourier analysis, (b) *Helping to clarify the structure of the new quantum mechanics* that emerged to replace the old quantum mechanics of 1900-1924 after the fundamental discoveries of Heisenberg and Schroedinger in late 1924 and early 1925; (c) Unifying spectral theory with the theory of group representations while applying both to the new quantum mechanics. (Mackey 1988, p. 138; emphasis added)

In the first edition of the book (1928) Weyl develops Heisenberg-Born-Jordan's formalism of Matrix Mechanics. However, in the second edition Weyl accepts a compromise (§. 2); he declares that de Broglie's and Schroedinger's "approach seems to me less cogent, but it leads more quickly to the fundamental principles of QM". (p. 48).

From the outset he made use of rather than real numbers, the algebra of specific fields of numbers which are suitable for the new TP ("Preface", p. viii), maybe the field of his "elementary mathematics" (Drago 2000a, Drago 2000b).

Weyl devotes Ch. 1 of his book to the notion of an affine space, viewed as the new mathematical framework for advanced TP. Physical states receive the Hermitian representation; operators are represented by matrices. The book extensively introduces into QM the group theory, permutation group too (for more details see Drago 2000b, pp. 401-403). But Weyl naively generalizes through mere analogies finite groups properties to those of infinity groups (Weyl 1928, pp. 28, 31, 34, 35; notice the use of Dirac's function in pp. 36-37).

> The heart of the book lies in the chapters II, IV and V. Chapter II contains one of earliest systematic coherent accounts of quantum mechanics as a whole. Perhaps only Dirac had as complete an overall view earlier but his early accounts are less complete and well organized. (Mackey 1988, p 147)

In synthesis, Weyl's presentation of QM includes three basic items, i.e. the representation of physical magnitudes in affine space, the mix of discrete-continuum in group theory, the derivation of the commutation relations. As its main result, the book suggested great and important applications of groups to QM, which however was no more than a great effort for founding QM (and special relativity) on group theory (ingenuously improved). His effort to represent a formulation of QM was unsuccessful, otherwise it would have been in the history the first physical theory based on groups ("Weyl Program"):

> While more suggestive than persuasive or logically compelling, [Weyl's results] are of importance as perhaps the first step in the program of deriving fundamental relationships in quantum mechanics from group theoretical symmetry principles in a program which I feel it appropriate to call Weyl program and to distinguish fairly sharply from the related important program inaugurated and much developed by Wigner.(Mackey 1988, p. 146)

Then, his attempt was also stopped by both Janos von Neumann theorems on the limitations of the method of matrices (Bernkopf 1968, pp. 341-346). Apparently as an answer to this stop, Weyl tried to improve his group theory (Weyl 1939) but Emmy Noether's contemporary axiomatization of group theory (a work considered by most mathematicians as the culmination of a half-hundred-year quest for the legitimization of algebra as a truly general, symbolic mathematics) de-evaluated his attempt as a backwards one.(Weyl 1939, p. vi) Again a very promising algebraic approach to QM was put aside by the developments of the contemporary scientific research.

## 4. Dirac's foundations (1930)

Dirac's celebrated book[1] locates QM inside a vector space (and its dual space) for representing both states as vectors and observables as matrices. Its basic mathematical technique is that of the "transformations" (actually, group theory) which later the reader recognizes to be the canonical transformation. After a first chapter of introduction to the principle of superposition as the main distinction in the description of quantum systems from the classical ones, five chapters follow about the mathematical preliminaries concerning algebraic structures and contact and canonical transformations. Dirac elaborated his algebraic framework at a so high degree of clarity and effectiveness to be qualified by him as a "symbolic method" (in the 3$^{rd}$ edition it was improved by the introduction of bra-ket formalism) and it greatly impressed its readers. Yet, to algebraic mathematics he added an idealistic function, the well-known Dirac's *$\delta(x)$*; therefore his foundation is not rigorously algebraic.

Moreover, the mathematical clothes of the theory cannot neither circumvent the basic analogy, nor justify it; indeed, it is an analogy which, although incomplete, allowed him to 1) suggest an important connection of QM with the more advanced classical mechanics (Hamiltonian formulation): the invariants to the transformations, i.e. Poisson brackets are equated to the commutators. 2) construct great part of previous QM (the indeterminacy principle, Matrix mechanics, Schroedinger representation and its equivalence with the previous one, electron dynamics (but not interactions), etc.); 3) add important theoretical improvements.

Yet, next editions of the book had to take in account some novelties: von Neumann's theorems which barred the way to a rigorous treatment of continuous variables in his matrix formalism, the criticism of a not rigorous treatment of the subject, and mainly the birth in the year 1932 of von Neumann's impressive formulation of QM through the axiomatic method. Dirac introduced some linkages to the new advancements, but mainly he declared to axiomatize his theory (4$^{th}$ edition, pp. 14-15), although in the first edition he had charged this method to be "somewhat artificial" and had declared to reject it (pp. 16-17).

This non linear evolution of his theory has justified the now usual appraisal of a first non rigorous introduction to QM, then correctly founded by von Neumann in the Hilbert space. This appraisal started with an influential von Neumann's remark:

> In the preface to his book on the foundations of quantum mechanics, von Neumann says of Dirac's own formulation of quantum theory that it is "scarcely to be surpassed in brevity and elegance," but that it "in no way satisfies the requirements of mathematical rigour." […] Since [von Neumann's] work was published, little has changed to affect the validity of these remarks [….] the Dirac formalism remains far from rigorous, and the formulation in terms of Hilbert space is still the only adequate framework for quantum theory. The very elegance and success of the Dirac formalism have ensured its survival. Most of the current generation of books on quantum theory prefer to take it as their guide, rather than give more than a passing reference to the niceties of Hilbert space. (Roberts 1966, p. 1097)

But the same Roberts stresses that the assimilation of the former book to the latter one is not correct:

> The most unsatisfactory feature of the present situation is that the gulf between the Dirac formalism and Hilbert space [perceived by those taking Dirac's book as a guide] is quite substantial, so that a lot of rethinking is necessary [to these scholars] before grasping the "correct" way of expressing things in Hilbert space. (*ibidem*)

---

[1] I will refer to the first edition of the book, which is online: https://archive.org/details/in.ernet.dli.2015.177580/page/n27/mode/2up. The 2$^{nd}$ edition is the same plus the table of content.

## 5. von Neumann's analytical formulation (1932) and his subsequent 'conversion' to an algebraic approach

The theoretical approach of the celebrated book (von Neumann 1932) is radically different from Dirac's approach because it was based on a separable infinite dimensional Hilbert space; shortly, analytical mathematics.

Von Neumann criticized Dirac's introduction of mathematical fictions, as he considered to be both $\delta(x)$ function and the assumption that any self-adjoint operator can be put in a diagonal form. He wanted a "just clear and unified [treatment of QM as Dirac's was] but without mathematical objections".

The great success of this book seemed to represent a decisive victory of the latter mathematical attitude, a confirmation of the analytical mathematical-physics of the previous century. Current undergraduate didactic presents in such a way QM: the only formulation of QM whose mathematics is formally valid.

But, just few years after this book, von Neumann often expressed his unsatisfaction for Hilbert space. The most known and remarkable one occasion is a 1935 letter to Birkhoff:

> I would like to make a confession which may seem immoral: I do not believe absolutely in Hilbert space any more. After all, Hilbert space (as far as quantum mechanical things are concerned) was obtained by generalizing Euclidean space, footing on the principle of 'conserving the validity of all formal rules' [...]. Now we begin to believe that it is not the vectors which matter, but the lattice of all linear (closed) subspaces. Because: 1) The vectors ought to represent the physical states, but they do it redundantly, up to a complex factor, only 2) and besides, the states are merely a derived notion, the primitive (phenomenologically given) notion being the qualities which correspond to the linear closed subspaces.(von Neumann 2005)

The reasons why von Neumann criticized and abandoned Hilbert space QM can be summarized as follows:
- Interpretational inconsistencies in Hilbert space QM

*No von Mises type relative frequency interpretation of quantum probability can be given*
- Presence of operationally meaningless mathematical operations and entities in Hilbert space QM

*Usual (composition) product of observables, phase in state vector*
- Mathematical pathologies in Hilbert space QM

*All selfadjoint unbounded operators do not form an algebra*
- New mathematical discoveries

*Existence of finite, continuous von Neumann algebras*
- Perceived greater conceptual coherence of operator algebraic QM

*Unique a priori quantum probability determined by quantum logic*

Again it is most remarkable that von Neumann's move beyond Hilbert space QM was *not* motivated by any of the following:
l. Empirical inadequacy of Hilbert space QM
2. Any new empirical/physical discovery
3. Mathematical imprecision/nonsense in Hilbert space QM. (Rédei 2002, pp. 240-241)

A scholar commented the above 'confession' as follows:
> the above words are often quoted as evidence that by 1935 von Neumann favoured the so-called 'lattice theoretical' (also called 'quantum logic') approach, in which the lattice P(H) of projections of infinite dimensional Hilbert space H, the 'logic' of quantum system, is considered as the basic object of the theory. (Rédei 1997, pp. 493-494)

Not before one century ago physicists have reflected on the mathematical nature of the set of measurement instrumentations at their disposal. First the paper (von Murray von Neumann 1936) showed that one can consider this set as an algebra of operators provided that their values are bounded; it was baptized "ring of operators", presently called "von Neumann algebra". In next years von Neumann's mathematical research tried to supersede Hilbert space by uniting probability and logic - since both are based on the concept of orthogonality - by means of continuous geometries with a transition probability. Physically, this choice is tantamount to considering the quantum theory with infinitely many particles as more fundamental than simple quantum mechanics. (Stoeltzner 2001, p. 53) His 'type 1 infinite' algebra (a particular ring of operators) corresponds to the infinitely dimensional separable Hilbert space that provides the rigorous framework for wave and matrix mechanics. In addition he investigated what today are called 'type II and type III von Neumann algebras'. However, his researches were unsuccessful because the continuous geometries did not directly generalize, as hoped by von Neumann, Hilbert space; their class is too broad for the purposes of axiomatizing QM.(Rédei 1997, p. 509)[2]

## 6. Segal's formulation of QM through a C*-algebra

In the year 1947 Irvin Segal obtained "spectacular results" (Jammer 1974, p. 381; this historian offered a clever, quick summary of Segal's papers; a critical illustration of Segal's results is given by Emch 1984, chp. 9, sect. 1). By essentially adding to a von Neumann's ring of operators a norm (which essentially is a bound; more precisely, it is a function which assigns a strictly positive length or size to each vector in a vector space), Segal (1947) introduced a C*-algebra (a Banach algebra together with an involution satisfying the properties of the adjoint). A C*-algebra of operators substantiates the relationship of QM with mathematics in a more general way than Heisenberg's.

He constructed a new axiomatic of QM on the base of a set of algebraic and metric axioms which define a more general structure than a C*-algebra (because the former one does not require to define adjoint operators; however, most followers dealing with QM exploited a simple C*-algebra). Segal's mathematics may seem less powerful than Hilbert space's one, which makes use of whatsoever (square summable) mathematical functions. However, more general functions may be obtained, so that a C*-algebra recovers Hilbert space. By completing other works on the representations of algebras, he showed that a C*-algebra can be represented within Hilbert space (Gelfand-Naimark-Segal's theorem; and *viceversa* the Stone-von Neumann theorem assures the equivalence of Hilbert space with C*-algebra in the finite case). Hence, here algebra comes first, then the the geometry of the representation into Hilbert space follows. As Segal himself remarks:

> it is interesting to note that if our algebraic postulates are strengthened sufficiently, then it can be shown that the collection of observables is isomorphic, (algebraically and metrically) with all self-adjoint operators in an algebra of bounded operators on Hilbert space (the norm corresponding to the operator bound). (Segal 1947, p. 931)

It was the first time that an entire formulation of QM was formulated in an algebraic way.

## 7. Jordan's didactic version of QM through matrices (1986)

H-B-J's algebraic approach to QM – i.e. to represent the measurable quantities of a quantum system through matrices - has been elaborated by a scholar, Thomas F. Jordan, in order to suggest an interesting didactic formulation of QM (Jordan 1986).

Since in general two matrices do not commute, as the physical variables of QM do, Jordan represents all physical magnitudes by means of finite matrices. He chooses as basic the system of the most simple matrices, Pauli's. His further choice of bounding the theory to finite or discrete variables dramatically simplifies the mathematical problems to be addressed so that "there is no calculus" in his proposal. The measurement is not a projection of an amplitude of probability for obtaining a probability result, but it is directly a probability as the mean of the values of the matrix representing the physical variable.

He starts by putting as the main problem of QM to explain "The Strange Equation", i.e. the relation of non commutation of the two conjugate variables, position and momentum. After a list of preliminaries, it is in chp. 14 that he easily obtains the states of a system of two non-commuting magnitudes; they represent through only Pauli's matrices two particles with spin 1/2. In chp. 17 Jordan obtains the principle of uncertainty of Heisenberg from the properties of matrix product: if A and B are matrices that have real eigenvalues, then the mathematical properties of matrices give the following equality:
$$\sqrt{\langle(A-\langle A\rangle)^2\rangle}\sqrt{\langle(B-\langle B\rangle)^2\rangle} = \tfrac{1}{2}|AB-BA|$$
where the symbol < > represents the average value and | | the module.

---
[2] A side product of this algebraic program of research was the discovery of quantum logic by him and Birkhoff (1936)- This logic is based on the modularity law (which replaces the distributivity law, failing in QM). Also this von Neumann's direction of research did not give the hope result of the true quantum logic. In the past eighty years and more this research (although corrected by replacing modularity with orthomodularity) is still ineffective; it was called "the quantum logic labyrinth".(van Frassen 1974)

Then, the consequences of this equation are derived. In chp.s 18-22 he finds out through the use of infinite matrices harmonic oscillator, Bohr's model, the quantized levels of angular momentum, rotational energy and at last hydrogen atom. Last chapters of Jordan's book deal with symmetries by treating only small continuous transformations. The expansion in series is considered and the generators are obtained; symmetry matrices are both physical variables and transformation generators. Of course, also these generators undergo the corresponding commutation relations. Then the commutation properties of all the physical quantities arise from the commutation properties of all the different kinds of transformations. In chp.s 23-27, spin rotation, small rotations, and, provided that continuum is bounded, changes in location, time and velocity; thus covering all successful cases of Heisenberg's theory (Beller 1989, p. 487).

Yet, the mathematical formalism is not sufficient for treating the entire QM, mainly because the continuous variables are bounded and the behavior of bosons (electromagnetic field) is ignored. Moreover he illustrates the dynamics by merely assuming the Hamiltonian as the equation of motion, and then he suggests its solutions to be merely verified as exact solutions.

In sum, his strictly algebraic theory offers a very simplified treatment of a substantial part of QM.

**8. Strocchi's formulation and his new quantization**
In 2012 Strocchi revisited "Dirac-von Neumann axioms of QM" and suggested a new formulation of QM as based on a C*-algebra inasmuch as he offered a new version of the last axiom:

> *Axiom A*. The observables generate a [polynomial] C*-algebra $\mathcal{A}$, with identity; the states which by eq. (2.1) define positive linear functionals on the Algebras $A_A \in \mathcal{A}$, for any observable A, separate such algebras in the sense of eq. (2.6) and extend to positive linear functional on $\mathcal{A}$.(Strocchi 2012, p. 6).

I illustrated in details this novelty - as well as previous case of Segal's formulation - in the papers (Drago 2018, sections 4-7).

Through algebraic methods Strocchi obtained an even more important result: an accurate definition of Dirac quantization; the original analogy was superseded by an exact set of algebraic rules(Strocchi 2018, chp. 7; Drago 2020); of which Dirac's identity
$$[A, C]\{B, D\} = \{A, C\}[B, D]; \qquad \forall A, B, C, D \in \mathcal{A};$$
is stressed and the crucial two

> 5) *Z relates the commutator to the Lie product, in the sense that* $\forall E, F, G, H \in \mathcal{A}$
> $$[E, F] = Z\{E, F\}, \quad [Z, \{G, H\}] = 0 = [Z, [G, H]].$$
> 6) *Z commutes with all the elements of $\mathcal{A}$, i.e. it is a central variable:*
> $$[Z, A] = 0 \qquad \forall A \in \mathcal{A}.$$

are justified. This result gave solution to an old problem about which along eighty years scholars hard worked.(Ali & Englis 2005) As an astonishing consequence, the two cases of classical (Hamiltonian) mechanics and QM correspond to two values (respectively 0 and $ih/2\pi$) of a parameter $\Lambda$ (I already illustrated this result in Drago 2020) This purely algebraic result frees the foundation of QM from both all past tentative analogies and the rough limit process $h \to 0$, which actually is a singular limit (Rohrlich 1990), hence insufficient to give great part of the mathematics of QM.

**9. Buchholz and Fredenhagen's formulation of QM**
Buchholz and Fredenhagen belong to the school of Loop Quantum Gravity, which joins these two physical theories by conceiving space as quantized by loops at the scale of $10^{-35}$m; these loops have energy but not in a unbounded quantity. Their paper (Buchholz and Fredenhagen 2020a) presents an approach to QM which is entirely based on concepts and facts taken from a classical world - i.e. classical mechanics whose dynamics is described by a Lagrangian - without imposing from the outset any quantization condition for observables. Quantum effects are inherited from the macroscopic arrow of time, determining the causal structure of operations; specific properties, such as commutation relations, then follow from the given dynamics.

The authors start from the configuration space of a finite number of $N$ particles (all of the same mass = 1); it is defined in the classical way, as $R^{sN}$, so that the state of the system is a vector $x$. Particles motions are orbits $c$; they are are described by functions (depending on time) of $x$ and forming a set $C$. In space $R^{sN}$ the basic mathematical structure will be defined as an algebra. This algebra describes our interventions into the quantum world. Such perturbations of the system typically arise from measurement arrangements, where forces can be manipulated in a systematic manner by human interventions within laboratories; but the surrounding of the system produces perturbations as well. They are described by functionals $F$ involving arbitrary potentials and some information as to when and for how long these perturbations act.

$$F[x] = \int dt F(x(t)), \quad \text{where } F((x(t)) = f_o(t)x(t) + Sg_k(t)V_k(x(t)),$$

where $f_o(x)$ is a fixed loop (a path starting and ending at 0); $g_k(t)$ is a test function, i.e. a smooth (that is infinitely often differentiable) function with compact support; a compact support for a function on the real line means that the function is zero outside of some finite interval; and $V_k(x(t))$ is a continuous bounded function describing a perturbation. Moreover, these functional can be shifted by loops $x_o \, \varepsilon \, C_o$, a shift is given by $F^{xo}[x] = F[x + x_o]$. Notice that the functionals $F$ describe the envisaged perturbations of the quantum system in "common language" of macroscopic world.

These functionals $F$ are non-linear, but enjoy the property which is manifest when dividing the full time axis in three disjoint pieces, representing the support of $x_1$, the support of $x_2$ and their common complement $x_3$.

$$F[x_1+x_2+x_3] = F[x_1+x_3] – F[x_3] + F[x_2+x_3].$$

The dynamics of the motions is governed by a *classical Lagrangian L* and the corresponding action, as it is familiar from classical mechanics.

The second ingredient taken from the macroscopic world is the arrow of time. Time enters into the quantum world since we can firmly state whether some operation was performed after or before some other operation. Moreover, it is impossible to make up for missed operations in the past, because time is directed. Since operations can be performed time and again, it is meaningful to assume that they form a semigroup. As a matter of fact, dealing with experimentally accessible systems, it is also plausible that the total effect of two successive perturbations, described by the sum of the underlying functionals, is equal to that of the product of the two individual perturbations. In addition, the effect of some operation can be rubbed out by another one, in accordance with the fact that experiments can be repeated and that their inverses represent the idea that in finite systems it is possible to remove the effects of a perturbation by other suitable perturbations. Therefore the operations, which can be performed on quantum systems, form *a group*, whose elements $S(F)$ are labeled by symbols on the functionals $F$, describing the perturbations. $S(F_1) \, S(F_2)$

Then, the elements $S(F)$ of the group satisfy two relations which describes *the dynamical evolution* of the system; the former one descries the action on the system descried by the Lagrangian:

1) $S(F) = S(F(x_o) + \delta L(x_o))$, for all $x_o \, \varepsilon \, C_o$, $F \, \varepsilon \, \mathcal{F}$;

and the second one translates previous temporal property of Functionals in algebraic terms:

2) $S(F_1+F_2+F_3) = S(F_1+F_2)S(F_3)^{-1} S(F_2 +F_3)$, for arbitrary functional $F_3$, provided $F_1$ lies in the future of $F_2$.

The second property is a *"causal" relation* imposing the direction of time because describes the ordering effects of time on the base of the fact that any functional comprises information as to when the corresponding perturbation takes place. This allows us to incorporate the arrow of time into the group by relying on the temporal order of perturbations. At this point the direction of time enters also at the microscopic level, because one can firmly state that some operation has happened earlier, respectively later, than another one. Owing to the directionality of time this group is non-commutative. So one could argue that it is this discrete arrow of time which is at the origin of the "quantization" of the classical theory. The specific form of commutation relations then follows from the underlying classical dynamics.(Buchholz and Fredenhagen 2020b)

In sum, the two basic ingredients, Lagrangian and time causality, determine the structure of the *dynamical group* of a Langrangian, $L$, $G_L$, whose element is designated by the generating symbol $S(F)$ and whose aim is to describe the dynamical effects of perturbations on the underlying system. Hence, without stipulating from the out-set their "quantization", their concrete implementation in the quantum world emerges from the inherent structure of the algebra.composed by Lagrangian and time causality.

Then, a standard procedure leads from the group $G_L$ to a C*-algebra $A_L$. This algebra is by definition the complex linear span of the elements $S \in G_L$. By easily equipping by a norm this algebra one obtains Segal's C*-algebraic setting.

The resulting C*-algebraic setting covers the full set of operations and resultant observables of the system. The C*-algebra contains in the non-interacting case unitary exponentials of the position and momentum operators (Weyl operators), satisfying the Heisenberg relations for position and velocity measurements. It is shown that Hilbert space representations of the algebra lead to the conventional formalism of quantum mechanics, where operations on states are described by time-ordered exponentials of interaction potentials. It is then proven that the dynamical algebras are irreducibly and regularly represented in the Schrödinger representation by time-ordered exponentials of functions of the position and momentum operators, described in terms of the classical theory. Interacting systems can be described within the algebraic setting by a rigorous version of the interaction picture. Thus, a "classical approach" reproduces the structure of QM in every respect.

## 10. Appraisals on the above foundations of QM according to the two basic dichotomies

In previous paper I suggested that the foundations of a scientific theory are constituted by the choices on two basic dichotomies, one on the kind of infinity (either potential (PI) or actual (AI) - in correspondence: either constructive mathematics or classical mathematics) and another on the kind of organization (either aimed at solving a problem (PO) or based on principles-axioms (AO) - in correspondence: making use of either intuitionist logic or classical logic).(Drago 2012)

The first dichotomy on the kind of mathematics is more general than the dichotomy algebra/analysis inside a specific kind of mathematics.

The problem I tackle now is which choices on the two dichotomies are at the base of the previous formulations.

*Heisenberg-Born-Jordan's formulation* makes use of finite matrices which manifestly belongs to a **PI** mathematics (apart the problem of the multiplicity of the eigenvalues of a matrix). Moreover, it is a **PO** theory inasmuch as it is aimed at solving the problem of extenting Bohr's atom theory.

*Weyl's book*. Its organization is not linear. It starts by merely describing the theoretical results of QM and then it applies group theory to only quantum general problems; hence, on the kind of organization there is no manifest choice. The book starts by dealing with finite groups (hence PI mathematics), but then generalizes without a support the results to infinite groups. Hence also on the kind of infinity there is no clear choice. (Drago 2000a)

*Dirac's* (first edition of the) book leads a reader to think that a basic mathematics solves all problems. His theory therefore relies on the algebra of vector space and group of transformations but it then applies the function $\delta(x)$), clearly appealing to **AI**. The mathematical clothes of the theory cannot cancel the crucial step of the development of the theory, which is an analogy aimed at overcoming g the radical difference between classical mechanics and QM. Hence, its choice is **PO** (and as a matter of fact, the theory is also based on the characteristic features of a PO theory; doubly negated propositions, *ad absurdum* proofs and the principle of sufficient reason).(Drago 2019, App.). (However, in the subsequent editions he aligned himself with the axiomatic approach; see pp. 14-15 of the fourth edition).

*Von Neumann's* search according to his new algebraic approach is manifestly based on the same choices of his book: **AI** and **AO**. (Drago 1991)

*Segal's* formulation is based on the choices **AI** (its two results have no counter-parts in constructive mathematics: a norm and Gelfand-Naimark-Segal's theorem (GNS) assuring a representation of a C*-algebra into Hilbert space; they) and **AO** (it suggests a set of axioms for QM). (Drago 2018)

*Jordan's didactic formulation* is manifestly based on the choice **PI** and may be presented in an even more clear **PO**. (Drago 2016, sect. 3)

*Morchio and Strocchi's "operative" formulation* of QM in C*-algebra relies on the choices **AI** and **AO**. Rather, Strocchi's rigorous translation of Dirac's quantization may be represented as a **PO** theory.(Drago 2020)

*Buchholz and Fredenhagen* apply to a discrete quantization a sophisticated mathematics concerning loops, test functions, the notion of a compact support. All together they represent a passage from the discrete loops to the continuum through a mere formula of an integral, which therefore appeals to **AI**. Owing to the a priori hypothesis of loops, the kind of organization is **AO**.

## 9. A table summarizing the eight algebraic approaches

| | *Basic physical notion* | *Basic physical principle* | *Basic math. notion* | *Mathem. technique* | *Result* | *Kind of infinity* | *Kind of organization.* | *Problems for MoC* |
|---|---|---|---|---|---|---|---|---|
| **Heisenberg-Born-Jordan (1925)** | Observables | Operativism, Non-commutation | Matrices | Algebraic calculus | Hydrogen atom | PI | PO | No (multiple eigenvalues?) |
| **Weyl (1928)** | System's state | Schroedinger Equation | Affine space | Transf. groups | Heisenberg Relations, Groups in QM | AI? | PO? | H |
| **Dirac (1930)** | System's state | Hamilton equations | Vector space | Transf. groups | Analogy with Hamiltonian | AI | PO | Imp ($\delta(x)$) |
| **(Murray &) von Neumann (1936)** | Linear closed subspace | Non-commutation | Ring | Continuous geometry | Type II and III algebras | AI | AO | Imp? |
| **Segal (1947)** | Physical operator | Non-commutation | C*-algebra | C*-algebra of operators, GNS | Axiomatic | AI | AO | H (GNS, norm) |
| **T.F. Jordan (1986)** | Spin | Hamiltonian | Pauli matrices | Algebra of finite matrices | Introduction to part of QM | PI | PO | No |
| **Strocchi (2012)** | Physical operators | Non-commutation | C*-algebra of operators | GNS representation in Hilbert space | "Operative" axiomatic | AI | AO | H (GNS, norm) |
| **Buchholz and Fredenhagen** | Perturbations | Lagrangian | Space of configurations, Loops | Group and C*-algebra of perturbatio | Algebraic foundation of QM | AI | AO | Imp? |

| (2020a) | | | | ns | | | | |

The column 9 presents evaluations of the difficulties met in translating the formulation into Constructive mathematics: "No" means no problem; "H" a hard goal and "Imp" a hopeless goal. Previous my communications to SISFA Congresses already presented these evaluations in a detailed way, except for the last foundation.

Let us read the table by columns (fourth and sixth rows are excluded because incomplete foundations), spanning over a century. The mathematical viewpoint is considered in columns 3 and 4; they shows a progress in introducing an ever more improved algebra structures for formulating QM; they range from Heisenberg's elementary technique – through vector space, group theory and C*-algebra - to Buchholz and Fredenhagen's framework.

Let us now consider the viewpoint of theoretical physics. The column 1 concerning the basic physical notion, manifests an exploiting of every aspect of a quantum system: observables, system's state, physical operators, perturbations. The column 2, about the basic principle of a foundation, manifests a tension to abandon the specific principle of QM, the non commutativity one, for recovering the classical framework of theoretical physics - Dirac and Strocchi the Hamilhonian one, Buchholz and Fredenhagen the Lagrangian one -, but now qualified as a universal theoretical framework.

Let us consider the columns 7 and 8 concerning the basic choices on the two dichotomies, actually ignored by all theoretical physicists. Four out eight foundations rely on the dominant choices AI&AO; at most two foundations (Dirac and maybe Weyl) rely on the mixed choices AI&PO. Only one foundation relies on the alternative choices PI and PO to the dominant ones; it is very important because it is the first foundation of QM (B-H-J), but it is algebraically incomplete; and it was not followed by other foundations based on the same choices. Hence, no historical progress occurred in exploring all the basic choices. In particualr, *a complete formulation of QM relying on the choices PI and PO has to be still discovered*.

## 11. Conclusions

Previous review showed that at last algebra plays a crucial role in founding QM foundations, so much to be put on the same par or even a more higher level of the analytical approach.

Current undergraduate didactics ignores the algebraic approach to QM because misinterpret Dirac's as an AO similar to the dominant one, in which albeit von Neumann himself distrust. The time has come for a radical reform of QM undergraduate didactic.

**Bibliography**
Ali S.T., Englis M. (2005). "Quantization Methods: A Guide for Physicists and Analysts". *Review of Mathematical Physics*, 17, pp. 391-490.
Beller M. (1989), "Matrix Theory Before Schroedinger. Philosophy, Problems, Consequences." *ISIS*, 74, pp. 469-491.
Bernkopf M. (1968), "A History of Infinite Matrices", *Arch. Hist. Ex. Sci.*, 4, 308-358.
Birkhoff G., von Neumann J. (1936). "The Logic of Quantum Mechanics", *Annals of Mathematics*, 37, pp. 823-843.
Bishop E. (1967), *Foundations of Constructive Analysis*, New York: AcademicP..
Born M., Heisenberg W., and Jordan P. (1925), "Zur Quantenmechanik II", *Zeitschrift für Physik*, **35**, 557-615, 1925 (B. L. van der Waerden (ed.) (1968), *Sources of Quantum Mechanics*, New York: Dover, pp. 321-386).
Buchholz D. and Fredenhagen K. (2020a), "Classical Dynamics, Arrow of Time, and Genesis af Heisenberg Commutation Relations", *Expositiones Mathematicae*, 38, pp. 150-167.
Buchholz D. and Fredenhagen K. (2020b), "From Path Integrals to Dynamic Algebras: A Macroscopic View of Quantum Physics", *Foundations of Physics*, 50 (7), pp. 727-734.
Darrigol O. (2014), *Physics and Necessity*, Oxford: Oxford U.P.

Dirac P.A.M. (1930), *Principles of Quantum Mechanics*, Oxford: Oxford U.P..

Donini E. (1983), "Meccanica ondulatoria e meccanica delle matrici: letture divergenti della meccanica classica", *Atti III Congr. Naz. Storia della Fisica*, Palermo, 1982, pp. 498-504.

Drago A. (1991), "Alle origini della meccanica quantistica: le sue opzioni fondamentali", in G. Cattaneo, A. Rossi (eds.): *I fondamenti della meccanica quantistica. Analisi storica e problemi aperti*, Cosenza: Editel, pp. 59-79.

Drago A. (2000a), "Which Kind of Mathematics for Quantum Mechanics? The Relevance of H. Weyl", in C. Garola, A. Rossi (eds.), *Foundations of Quantum Mechanics. Historical Analysis and Open Questions,* Singapore: World Scientific, 167-193.

Drago A. (2000b), "A New Mathematics for the New Theory of Quantum Mechanics. The Fate of H. Weyl's Program of Research", *Atti Fond. Ronchi*, 55 (2000), pp. 387-425.

Drago A. (2012). "Pluralism in Logic. The Square of Opposition, Leibniz's Principle and Markov's Principle". In J.-Y. Beziau, & D. Jacquette (Eds.), *Around and beyond the Square of Opposition* , Basel: Birckhaueser, pp. 175-189

Drago A. (2018), "Is the C*-Algebraic Approach to Quantum Mechanics an Alternative to the Dominant One?", *Advances in History of Science*, 7 (2).

Drago A. (2019), "A Mathematical Structure Defining the Incommensurability of Quantum Mechanics and Classical Mechanics", *Atti della "Fondazione Giorgio Ronchi"*, 74, n. 1, pp. 119-129.

Drago A. (2020), "Dirac's Quantization Improved by Morchio & Strocchi to an Algebraic Structure Founding Both Classical Mechanics and Quantum Mechanics", Esposito S. et al. (eds.), *Atti XXXVIII Congresso Nazionale della Società Italiana degli Storici della Fisica e dell'Astronomia Messina*, 2018, Pavia U.P., Pavia, pp. 65-73.

Emch G.G. (1984), *Mathematical and Conceptual Foundations of Quantum of $20^{th}$ Century Physics*, Amsterdam: North-Holland.

Giannetto E.A. (1997), "Note sulla rivoluzione della meccanica delle matrici di Heisenberg, Born e Jordan e sul problema dell'equivalenza con la meccanica di Schrödinger", in Tucci P. (ed.), *Atti del XVII Congresso Nazionale di Storia della Fisica e dell'Astronomia*. Milano: Università Milano, pp. 199-208.

Hanson N. (1963), *The Concept of the Positron. A Philosophical Analysis*, Cambridge University Press, Cambridge.

Heisenberg, W. (1927), "Über den anschaulichen Inhalt der quantentheoretischen Kinematik und Mechanik", *Zeitschrift für Physik*, **43** (3–4): 172–198.

Jordan T.F. (1986), *Quantum Mechanics in Simple Matrix Form*. New York: Wiley & Sons.

Mackey G.W. (1988), "Hermann Weyl and the Application of Group Theory to Quantum Mechanics", in Doppelt W. et al. (eds.) *Exacts sciences and Their Philosophical Foundations*, Frankfurt: Lang, pp. 131-159.

Markov, A. A. (1962). "On Constructive Mathematics". *Trudy Matematicieskie Institut Steklov, 67,* 8-14. [English Translation in (1971) *American Mathematical Society Translations, 98,* 1-9.]

Morchio G., Strocchi F. (2009). "Classical and Quantum Mechanics from the Universal Poisson-Rinehart Algebra of a Manifold", *Reports of Mathematical Physics*, 64, pp. 33-48.

Muller F. A. (1997), "The Equivalence Myth of Quantum Mechanics", I & II, *Studies in History and Philosophy of Modern Physics*, 28 , pp. 35-61, and.28, pp. 219-247.

Murray F.J., Von Neumann J. (1936), "On Rings of Operators", *Ann. of Math.*, 37, pp. 116–229.

Rédei M. (1997), "Why John von Neumann did not Like the Hilbert Space Formulation of Quantum Mechanics (and What he Liked instead)", *Stud. Hist. Phil. Modern Sci.*, 27, 4, pp. 493-510.

Rédei M. (2002), Mathematical Physics and Philosophy of Physics (with special consideration to von Neumann's work)", in Heidelberg M. and Stadler F. (eds.), *History of Philosophy of Science*, Dordrecht: Kluwer, pp. 239-243.

Segal I.E. (1947), "Postulates for General Quantum Mechanics", *Ann. Math.*, 48, pp. 930-948.

Stoeltzner M.(2001), "Opportunistic Axiomatic – von Neumann and the Methodology of Mathematical Physics", in Redéi M. and Stoeltzner M. (eds.), *John von Neumann and the Foudations of Physics*, Dordrecht: Kluwer, pp. 35-62.

Strocchi F. (2012), "The Physical Principles of Quantum Mechanics: A Critical Review". *The European Physical Journal Plus*, 127, 12.

Strocchi F. (2018), *A Primer of Analytical Mechanics*, Berlin: Springer, Chp. 7.

van Frassen B.C. (1974), "The Labyrinth of Quantum Logic", in Cohen R.S. (ed.), *Logical and Epistemological Studies in Contemporary Physics*, BSPS no. 13, Dordrecht: Reidel, pp. 224-254.

von Neumann J. (1932), *MathematischeGrundlagen der Quantenmechanik*, Springer, Berlin.

von Neumann J. (2005), *Selected Letters*, M. Rédei (ed.), Providence: American Mathematical Society.

Weyl H. (1928), *Group Theory and Quantum Mechanics*, New York: Dover.

Weyl H. (1939), *The Classical Groups*, Princeton: Princeton U.P.

.